\documentclass{aa}   

\usepackage{graphicx}
\usepackage{refcount}

\usepackage{txfonts,textcomp}
\usepackage{amsmath}  
\usepackage{gensymb}
\usepackage[normalem]{ulem}
\usepackage{comment}
\usepackage{array}
\usepackage{float}
\usepackage{lipsum}
\usepackage{adjustbox}

\usepackage{threeparttable}

\usepackage[colorlinks =TRUE, linkcolor = blue,
            urlcolor  = blue,
            citecolor = blue,
            anchorcolor = blue]{hyperref}

\begin{document} 

\title{X-ray polarization of the Compton-thin \\Seyfert galaxy NGC~5506 with IXPE}

\titlerunning{NGC~5506 polarimetry}
\author{Daniele Tagliacozzo\inst{1,}\inst{10}* \and
 Alessandro Leonardo Lai\inst{1} \and
 Andrea Gnarini\inst{8} \and
 Giorgio Matt\inst{1} \and
 Stefano Bianchi\inst{1} \and
 Francesco Ursini\inst{1} \and
 Andrea Marinucci\inst{4} \and
 Alessia Tortosa\inst{2} \and
 Sudip Chakraborty\inst{8} \and
 Federico Ferretti\inst{1} \and
 Vittoria Elvezia Gianolli\inst{3,}\inst{11} \and
 Elias Kammoun\inst{6} \and
 Frédéric Marin\inst{5} \and
 Herman Marshall\inst{7} \and
 Riccardo Middei\inst{2} \and
 Romana Mikusincova\inst{9} \and
 Dawoon Kim\inst{9}}
 \authorrunning{D. Tagliacozzo et al.}
   \institute{Dipartimento di Matematica e Fisica, Università degli Studi Roma Tre, via della Vasca Navale 84, 00146 Roma, Italy
             \and            
        INAF - Osservatorio astronomico di Roma, Via Frascati 33, I-00040 Monte Porzio Catone, Italy
        \and
        Dep. of Physics and Astronomy, Clemson University, Kinard Lab of Physics, 140 Delta Epsilon Ct, Clemson, SC 29634, USA
        \and
        Agenzia Spaziale Italiana, Via del Politecnico snc, 00133 Roma, Italy
        \and
        Universit\'e de Strasbourg, CNRS, Observatoire Astronomique de Strasbourg, UMR 7550, 11 rue de l'universit\'e, 67000 Strasbourg, France
        \and
        Cahill Center for Astronomy \& Astrophysics, California Institute of Technology, 1216 East California Boulevard, Pasadena, CA 91125, USA
        \and
        MIT Kavli Institute for Astrophysics and Space Research, Massachusetts Institute of Technology, 77 Massachusetts Avenue, Cambridge, MA 02139, USA
        \and
        NASA Marshall Space Flight Center, Huntsville, AL 35812, USA
        \and
        INAF Istituto di Astrofisica e Planetologia Spaziali, Via del Fosso del Cavaliere 100, 00133 Roma, Italy
        \and
        Istituto Nazionale di Fisica Nucleare, Sezione di Roma “Tor Vergata”,
via della Ricerca Scientifica 1, 00133 Roma, Italy
\and
INAF-Osservatorio Astronomico di Brera, Via Brera 28, 20121 Milano, Italy\\\\
        \email{*daniele.tagliacozzo@uniroma3.it}     }

   \date{Received Month Day, 2025; accepted Month Day, 2025}

   \abstract
{We report on the observation of the Compton-thin Seyfert galaxy NGC~5506 performed with the Imaging X-ray Polarimetry Explorer (IXPE). The observation, with a net observing time of 975~ks, was partly simultaneous with \textit{NuSTAR} (52~ks) and \textit{XMM-Newton} (37~ks). Using a simple baseline model with IXPE data only, we find an upper limit on the 2--8~keV polarization fraction at the $99\%$ confidence level: $\Pi<3.1\%$. We then performed a spectropolarimetric analysis combining data from the three X-ray observatories. While pure spectral modeling confirms the presence of both relativistic and cold distant reflection, in addition to the primary continuum power law,  their contributions lead to somewhat different constraints on the coronal $\Pi$, depending on the specific assumptions. We also performed Monte Carlo simulations with the radiative transfer code \texttt{MONK}, exploring various coronal geometries. Comparing these results with the IXPE data, and assuming a radially extended corona, as suggested by previous observations of similar sources, the corresponding polarization upper limits require inclination angles below $40\degree$.}


    \maketitle
 
\section{Introduction}
\label{Introduction}

Active galactic nuclei (AGNs) release vast amounts of energy, primarily originating from a compact central zone where matter accretes onto a supermassive black hole (SMBH; \citealt{Rees84, 1993ARA&A..31..473A}). 
There, matter forms into a rotating accretion disk around the central object.
This structure emits optical and UV radiation, a portion of which is subsequently up-scattered into X-rays through Comptonization in a cloud of hot, relativistic electrons: the X-ray corona \citep{1980A&A....86..121S, 1991ApJ...380L..51H, 2000ApJ...542..703Z}. 
The corona is characterized by electron temperatures, $kT_{\rm e}$, spanning from tens to hundreds of keV, along with moderate Thomson optical depths, $\tau$ (\citealt{torto2018A&A...614A..37T} and reference therein). 
In this context, deciphering the precise shape and structure of the AGN corona remains a significant challenge. 
In fact, while spectroscopic observations yield insights into the physical characteristics of the Comptonizing medium, they are insufficient to distinguish between different geometric configurations, which result in equally acceptable spectral fits (e.g., \citealt{dauser2012MNRAS.422.1914D, middei2019A&A...630A.131M}). 
In recent years, time lag measurements such as reverberation mapping, combined with disk emissivity profiles and reflection studies, have provided constraints on the location, shape, and size of the corona (e.g., \citealt{wilkins2012MNRAS.424.1284W, uttarticle, wilkinsgallo2015MNRAS.448..703W, wilkins2016MNRAS.458..200W, wilkins2017MNRAS.471.4436W, gonza2017MNRAS.472.1932G}). These studies also show that specific reflection models can reproduce the observed spectra \citep{nekra2025A&A...704A.129N}.
Nevertheless, many questions remain.
X-ray polarimetry provides a crucial observational approach. 
It offers a unique window into the geometry and physical properties of the corona because different structural configurations and compositional features are expected to produce distinct polarization signatures in the emitted radiation \citep{2022MNRAS.510.3674U, Tagliacozzo2025}.

Several models have been proposed for the coronal geometry. 
Each of them reproduces the spectral data equally well, but, as discussed above, they result in specific X-ray polarization properties. The spherical lamp-post, modeled as a simple isotropic cloud located on the spin axis of the SMBH, produces weakly polarized radiation ($\Pi=0-2\%$) with polarization angles ($\Psi$) perpendicular to the disk axis because of its very high degree of symmetry \citep{matt1991A&A...247...25M,marto1996MNRAS.282L..53M,pop1997A&A...326...99P,wilkins2012MNRAS.424.1284W,ursi2020A&A...644A.132U}. The conical outflow, associated with failed jets in radio-quiet AGNs, produces $\Pi$ up to $5-7\%$, with $\Psi$ also perpendicular to the disk \citep{conerefId0}. The slab corona is thought to arise from magnetic reconnection and is assumed to uniformly sandwich the disk. Because of its low degree of symmetry, it is expected to produce relatively high $\Pi$ (up to $14\%$), with $\Psi$ parallel to the disk axis \citep{1979ApJ...231L.111L, 1991ApJ...380L..51H, belo2017ApJ...850..141B}. Finally, the wedge geometry is similar to the slab but has a height that increases with radius.
It corresponds to a hot accretion flow that replaces the disk close to the SMBH.
It yields $\Pi$ up to $5-10\%$, with $\Psi$ also aligned with the disk axis (\citealt{Esin_1997, esin1998ApJ...505..854E, sh2010ApJ...712..908S, poutanen2018}).
Additionally, within a particular morphological configuration, variations in medium characteristics (such as $kT_e$ or $\tau$) can result in further differences in both $\Pi$ and $\Psi$ (see e.g., temperature differences in the wedge corona scenario in \citealt{Tagliacozzo2025} and differences in coronal size and primary continuum spectral index in \citealt{2022MNRAS.510.3674U}).
The polarized reflected emission also needs to be considered when interpreting X-ray polarimetric observations. 
\cite{dov2011ApJ...731...75D} showed that relativistic reflection from the accretion disk can produce a strong, energy-dependent polarization signal, particularly for compact lamp-post coronae. 
More recently, \cite{pod2023MNRAS.524.3853P} and \cite{pod2026A&A...711A.175P} investigated the combined polarization of the primary and reflected components, explicitly accounting for relativistic disk reflection. 
In particular, \cite{pod2026A&A...711A.175P} obtained total polarization fractions of up to $9\%$ for a lamp-post coronal geometry, highlighting the potentially significant contribution of reflection to the observed polarization signal.

The NASA and Italian Space Agency mission Imaging X-ray Polarimetry Explorer (IXPE; \citealt{weiss_ixpe_2022JATIS...8b6002W}), launched in 2021, is providing fundamental insight into the 2--8~keV polarization properties of radio-quiet, Compton-thin AGNs.
For these sources, this energy band is usually dominated by coronal emission, and this kind of information can help break the degeneracies between models.
During its first years of operations, IXPE observed MCG-05-23-16 (three times, in May and November 2022 and in April 2025; \citealt{Marinucci_2022, taglia2023MNRAS.525.4735T, 2026A&A...707A.154M}), IC~4329A (in January 2023; \citealt{ingram2023MNRAS.525.5437I}), NGC~4151 (twice, in December 2022 and in May 2024; \citealt{Gianolli2023, giano2024A&A...691A..29G}), NGC~2110 (in October 2024; \citealt{chakra2025ApJ...990...89C, 2025A&A...697A.182P}), and NGC~3227 (in May 2025; Chakraborty et al., in preparation):
NGC~4151 yielded a significant polarization detection in both observations. 
When most of the polarized signal is attributed to the primary emission, as suggested by the spectral decomposition, the polarization degree reaches $\Pi=(7.1\pm1.2)\%$, while $\Psi$ is consistent with the orientation of the radio jet, adopted as a proxy for the accretion disk axis.
This interpretation favors radially extended coronal geometries. 
However, \cite{Kammoun2026} show that the observed polarization can also be reproduced by an alternative scenario in which the primary coronal emission is intrinsically unpolarized and the signal is mainly produced by a highly polarized relativistic reflection component, with its polarization angle parallel to the accretion disk axis. For IC~4329A, IXPE measured polarization at a 2.97$\sigma$ confidence level, with $\Pi = 3.3\% \pm 1.1\%$.
The most probable value of $\Psi$ is consistent with the jet position angle, again favoring radially extended coronal models. In the case of MCG-05-23-16, the combined analysis of the three observations resulted only in an upper limit $\Pi < 2.5\%$ (at the $99\%$ confidence level) and, therefore, $\Psi$ is unconstrained. However, the contour plot suggests that an alignment between $\Psi$ and the position angle of the \textsc{[O~III]} emission has the highest probability, again hinting at a radially extended corona. For NGC~2110, a low-luminosity AGN, we obtain only an upper limit of $\Pi<10.1\%$, at the $99\%$ confidence level, which provides no useful constraints on the polarization direction. The analysis of NGC~3227 is still ongoing and the corresponding paper is in preparation.

In summary, IXPE results show that, owing to hints of alignment between the polarization angle of the coronal emission and the disk symmetry axis, radially extended coronal models are favored over vertically extended ones.
Nevertheless, the majority of the observed sources resulted in upper limits at the $99\%$ confidence level. In these cases, tight constraints on $\Psi$ are not possible.
Moreover, many results are model-dependent, allowing for different interpretations.
For these reasons, additional polarimetric observations are needed to resolve this issue.

In this context, IXPE observed NGC~5506, a nearby ($z=0.0062$) Compton-thin Seyfert galaxy that has been widely studied in X-rays. 
It does not show broad emission lines in the optical \citep{wilson10.1093/mnras/177.3.673}, but it exhibits broad lines in the IR, with velocities $v < 2000$ km s$^{-1}$ \citep{refId0}, and is moderately absorbed in X-rays ($N_{\rm H} \sim 3\times10^{22}$ cm$^2$; \citealt{bianchirefId0, guai10.1111/j.1365-2966.2010.16805.x}). 
It shows a relatively high 2--10~keV flux ($F_{2-10} = 5{-}10\times10^{-11}$ erg cm$^{-2}$ s$^{-1}$, dominated by primary continuum emission; \citealt{sun2018MNRAS.478.1900S}) and flux variability on short timescales without evidence of significant spectral variability \citep{bianchirefId0, matt10.1093/mnras/stu2653}. 
It is among the Seyferts with the highest lower limits on the primary continuum cutoff energy measured to date \citep{matt10.1093/mnras/stu2653}.
In soft X-rays ($<$2 keV), the spectrum is dominated by emission from diffuse gas associated with extended narrow-line regions \citep{sun2018MNRAS.478.1900S}.
Its SMBH mass is poorly constrained.
 \citep{Hayashida_1998} estimated $M_{\rm BH}=2.6\times10^6M_\odot$ from X-ray variability, while  \citep{guai10.1111/j.1365-2966.2010.16805.x} estimated a black hole mass of around $10^8M_\odot$ from the stellar velocity dispersion and \textsc{[O~III]} width.
Here, we use the relation found by \cite{torto2023MNRAS.526.1687T}, based on X-ray variability, to provide a new mass estimate. We used the 37 ks light curve obtained with \textit{XMM} to retrieve $M_{\rm BH}=(3.25\pm0.05)\times10^6M_\odot$.
This uncertainty strongly affects the corresponding Eddington ratio, which, considering $L_{2-10} = 5\times10^{42}$ erg s$^{-1}$ \citep{matt10.1093/mnras/stu2653}, ranges between 0.1$\%$ and 12$\%$. 
A relativistic reflection analysis reveals that NGC~5506 hosts a highly spinning SMBH, with $a = 0.93\pm0.04$, and has a moderately inclined accretion disk ($40\degree< \theta_{\rm disk} < 50\degree$; \citealt{sun2018MNRAS.478.1900S}). 
This result differs from the higher disk inclination reported by \cite{Fischer_2013}, who found $\theta_{\rm disk}\sim80\degree$). 
Moreover, from \textsc{[O~III]} imaging, \cite{Fischer_2013} found the narrow-line region biconical structure (assumed as a proxy of the disk orientation) to be in the $i\sim22\degree$ direction with respect to the north axis. 

In this paper, we present and discuss the IXPE observation of NGC~5506, performed in July 2025 in coordination with \textit{XMM-Newton} and \textit{NuSTAR}. 
We compare the results with dedicated Monte Carlo simulations of the expected polarization properties for different geometries of the corona.
The paper is organized as follows: Sect. \ref{red} discusses the data reduction procedure; Sect. \ref{analisi} presents the spectral and polarimetric analysis; Sect. \ref{sim}  presents dedicated Monte Carlo simulations; and, finally, we summarize the results in Sect. \ref{concl}.

\section{Observations and data reduction}
\label{red}
 In July 2025, IXPE observed NGC~5506, partly simultaneously with \textit{XMM-Newton} and \textit{NuSTAR}, for a net exposure time of 975~ks.
We produced and calibrated  the cleaned level 2 event files using standard filtering criteria, the dedicated \texttt{ftools} tasks, and the latest calibration files available in the IXPE calibration database (CALDB v.20250225). 
We extracted the $I$, $Q$, and $U$ Stokes background spectra from source-centered annular regions with an inner radius of 150~arcsec and an outer radius of 300~arcsec. 
We optimized the source extraction regions to maximize the signal-to-noise ratio (S/N) in the 2--8~keV band, following the procedure described by \citet{Piconcelli2004}, leading to the choice of a 75~arcsec radius for the first detector unit (DU) and 70~arcsec radii for the second and third DUs. 
We applied these same radii, centered on the source,  to the $I$, $Q$, and $U$ spectra. 
We used a constant energy binning of 0.2 keV for the $Q$ and $U$ Stokes spectra and required a S/N greater than 3 in each spectral channel, in the intensity spectra.
We always fit the $I$, $Q$, and $U$ Stokes spectra from the three DUs independently, but we plot them together using the \texttt{setpl group} command in \texttt{xspec}, for visual clarity. 
The background represents 2.7, 3.1, and 2.6 $\%$ of the total DU1, DU2, and DU3 $I$ spectra, respectively. 
We followed the formalism discussed in \citet{stro17} and used the weighted analysis method presented in \citet{dimarco22} (parameter \texttt{stokes=Neff} in \texttt{xselect}).  

The \textit{XMM-Newton}/EPIC pn \citep{xmm2001A&A...365L...1J} observed NGC~5506 for a net observing time of 37~ks. 
We processed the data using SAS v.21 \citep{Gabriel2004}, and adopted an iterative procedure to determine  the optimal source extraction radii and time cuts for background flaring, with the goal of maximizing the S/N, similar to the approach described by \cite{Piconcelli2004}. 
We extracted the background from circular regions with a radius of 50~arcsec, while we used an optimized source extraction radius of 40~arcsec.
We generated ancillary response files with \texttt{arfgen}, enabling \texttt{applyabsfluxcorr=yes} to improve the cross-calibration between \textit{XMM-Newton} and \textit{NuSTAR}.
After testing various spectral-binning procedures that yielded almost identical results, we applied an optimal spectral-binning procedure (\texttt{optsmin}) to the EPIC pn spectrum, following the algorithm of \cite{Kaastra2016} and ensuring a minimum S/N of 3 per bin.

The \textit{NuSTAR} telescope \citep{nustar2013ApJ...770..103H} observed NGC~5506 for a net observing time of 52~ks.
We reduced the data using the \texttt{nupipeline} task within NuSTARDAS v.2.1.4, using the most recent calibration database (CALDB v.20250428 at the time of analysis). 
We extracted spectra and light curves with \texttt{nuproducts} for both focal plane modules (FPMA and FPMB), following standard procedures. 
As for the IXPE data, we selected the source and background regions using an S/N-optimization procedure.
We extracted the background from circular regions with a radius of 100~arcsec, while the optimized source extraction radius was 120~arcsec.
For \textit{NuSTAR} data, we applied the same procedure as for \textit{XMM} for consistency, ensuring a minimum S/N ratio of 3 for each bin.

\begin{figure*}
\centering
\includegraphics[width=2.\columnwidth] {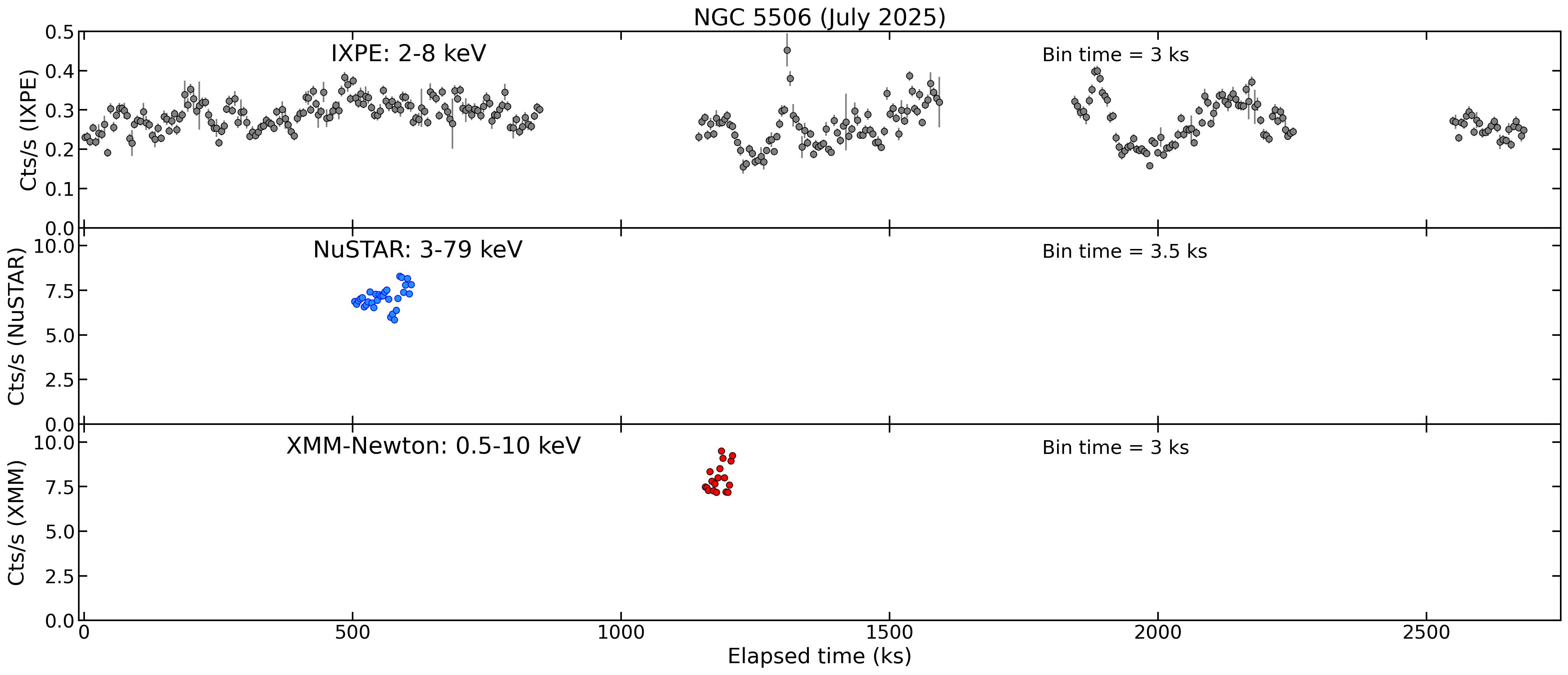}
\caption{Light curves of the NGC~5506 observing campaign obtained with IXPE, \textit{NuSTAR,} and \textit{XMM-Newton}. We summed the counts from DU1, DU2, and DU3 on board IXPE and from FPMA and FPMB on board \textit{NuSTAR}.  We used the optimized 2--8~keV energy band for IXPE (3~ks binning), 2--10~keV for \textit{XMM} (3~ks binning), and 3--79~keV for \textit{NuSTAR} (3.5~ks binning).}\label{lc}
\end{figure*}

We show the summed background-subtracted light curves for the IXPE, \textit{XMM-Newton}, and \textit{NuSTAR} pointings in Fig. \ref{lc}, with summed data counts from DU1, DU2 and DU3 on board IXPE and from FPMA and FPMB on board \textit{NuSTAR}.

In this work, we adopted the cosmological parameters $H_0=70$ km s$^{-1}$ Mpc$^{-1}$, $\Omega_\Lambda=0.73$, and $\Omega_{\rm m}=0.27$, that is, the default values in \texttt{xspec v.12.13.1e} \citep{1996ASPC..101...17A}. 

\section{Spectral and polarimetric analysis}
\label{analisi}

\subsection{IXPE analysis}
\label{ixpeanalysis}

We performed a preliminary evaluation of the polarization properties of the 2--8~keV data collected by IXPE.
To do this, we simultaneously fit the $I$, $Q$, and $U$ spectra from the three DUs over the 2--8~keV energy range, employing the following baseline model:\\\\  $\texttt{const} \times \texttt{tbabs} \times \texttt{polconst} \times \texttt{ztbabs} \times \texttt{powerlaw}$.\\\\
We used the multiplicative constant (\texttt{const}) to account for cross-calibration uncertainties between DU1, DU2, and DU3; \texttt{powerlaw} represents the primary continuum; and \texttt{polconst} is a constant polarization kernel.
Additionally, we used \texttt{tbabs} to account for Galactic absorption, assuming $N_{\mathrm{H,Gal}}=4.23\times10^{20}~\mathrm{cm}^{-2}$ \citep{HI4PI2016}, while we used \texttt{ztbabs} to model cold absorption at the redshift of the source, freezing it to the value derived from the broadband  spectral analysis presented in Sec. \ref{spectralanalysis} ($N_{H, z}=3.04\times10^{22}~\mathrm{cm}^{-2}$).
Previous studies have already provided evidence for this component \citep{bianchirefId0}. We verified its significance by performing fits with and without this component. We obtain a significant improvement in the reduced $\chi^2$, which decreases from $\chi^2$/degree of freedom (d.o.f.)=2.02 to $\chi^2$/d.o.f.=1.09.
This last result also demonstrates that more complex models are unnecessary when using only IXPE data.
From this fit, we find a primary continuum spectral index of $\Gamma=1.85_{-0.04}^{+0.03}$. 
We obtain a polarization degree $\Pi=(1.0\pm0.8)\%$ and a polarization angle $\Psi=38\degree\pm26\degree$ at the $68\%$ confidence level for one interesting parameter, which becomes an upper limit of  $\Pi<3.1\%$ at the $99\%$ confidence level.

\begin{figure*}
\includegraphics[width=8.8cm] {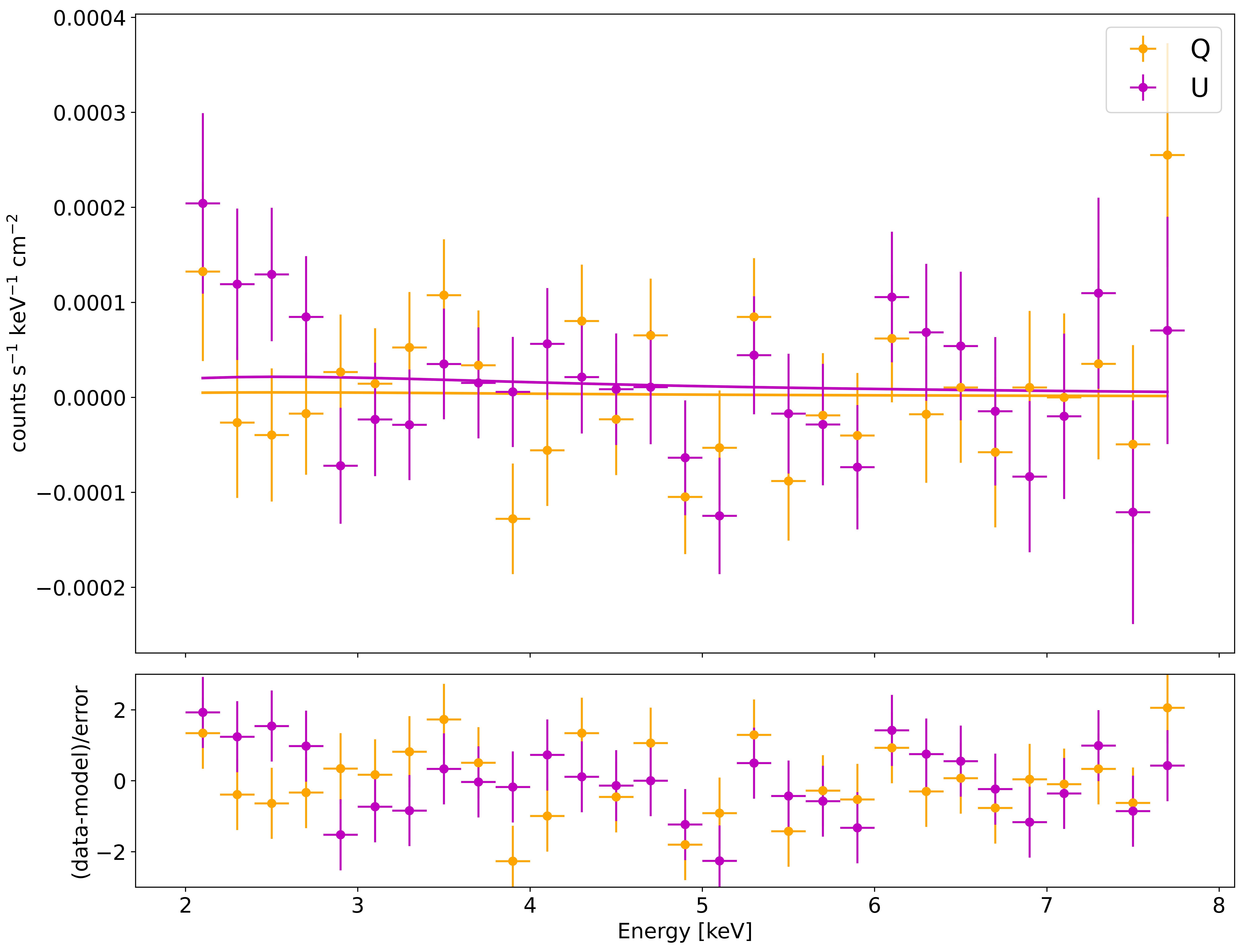}
\includegraphics[width=8.8cm] {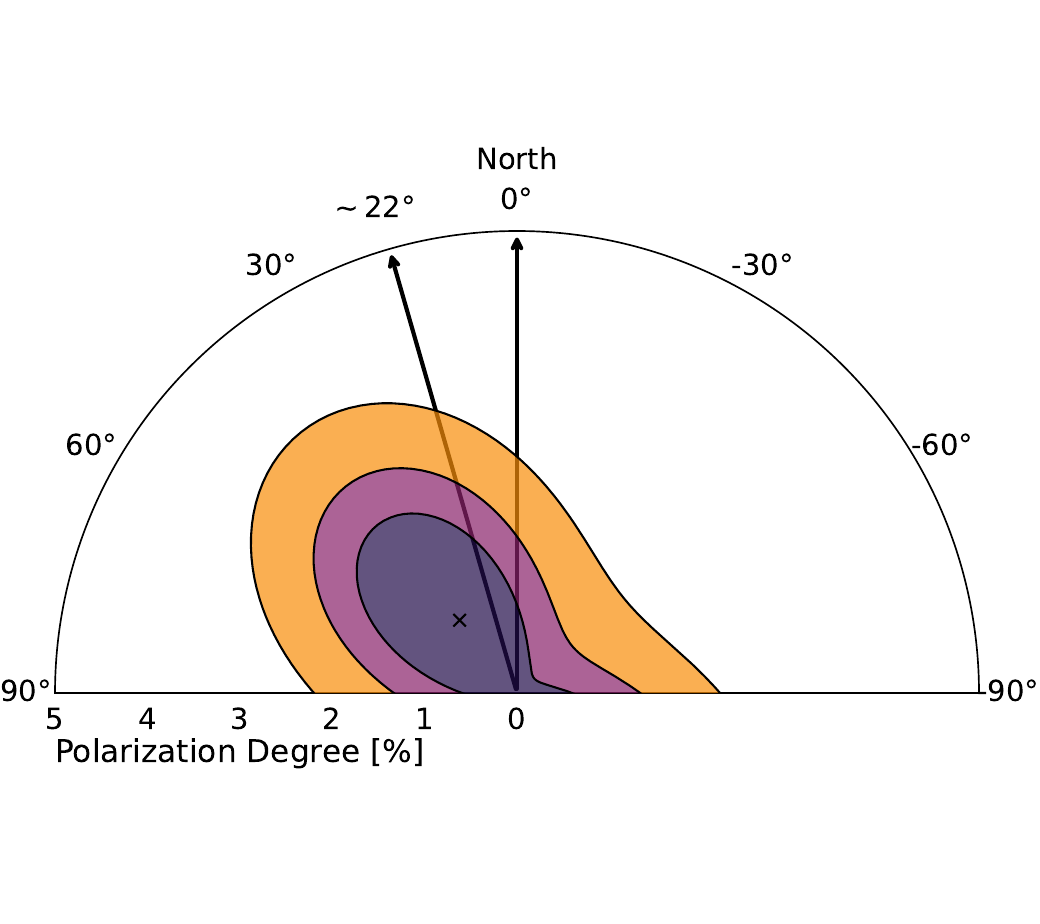}
\caption{\textit{Left:} Grouped IXPE $Q$ (orange crosses) and $U$ (magenta crosses) Stokes spectra of NGC~5506, together with the corresponding best-fitting model and residuals. \textit{Right:} Contour plot of the polarization degree $\Pi$ and angle $\Psi$, summed over the 2--8~keV energy band, for a model-independent analysis. Purple, pink, and orange regions correspond to the $68\%$, $90\%$, and $99\%$ confidence levels for the two parameters of interest. The solid black line at $0\degree$ represents the north direction, while the line at $\sim22\degree$ represents the orientation of the narrow-line region \citep{Fischer_2013}.}
\label{Fig:ixpemodindep}
\end{figure*}

The left panel of Fig. \ref{Fig:ixpemodindep} presents the $Q$ and $U$ spectra used in this initial analysis, including the model and residuals, while the right panel shows the contour plot between $\Pi$ and $\Psi$.
As in \cite{Gianolli2023, giano2024A&A...691A..29G}, we divided the dataset into three energy bins (i.e., 2--3.5~keV, 3.5--5~keV, and 5--8~keV), and searched for deviations in $\Psi$.
At the $99\%$ confidence level, we find only upper limits in each energy bin.
Thus, the analysis does not provide useful constraints on the polarization direction.

\subsection{broadband spectral analysis}
\label{spectralanalysis}
Next, to achieve a thorough understanding of the system and gather relevant parameters for an effective spectropolarimetric analysis, we conducted a broadband X-ray spectral analysis. 
This analysis combined the complete set of IXPE $I$ spectral data (DU1, DU2, and DU3; 2--8~keV) with the 3--79~keV \textit{NuSTAR} (FPMA and FPMB) and the 2--10~keV \textit{XMM-Newton} (EPIC pn) data.
To correct calibration issues in the IXPE data, we followed the procedure of \cite{Marinucci_2022} and \cite{taglia2023MNRAS.525.4735T}, adjusting the gain of the $I$ spectrum using the \texttt{gain fit} command in \texttt{xspec}.

\begin{figure*}[t]
\sidecaption
\includegraphics[width=12cm]{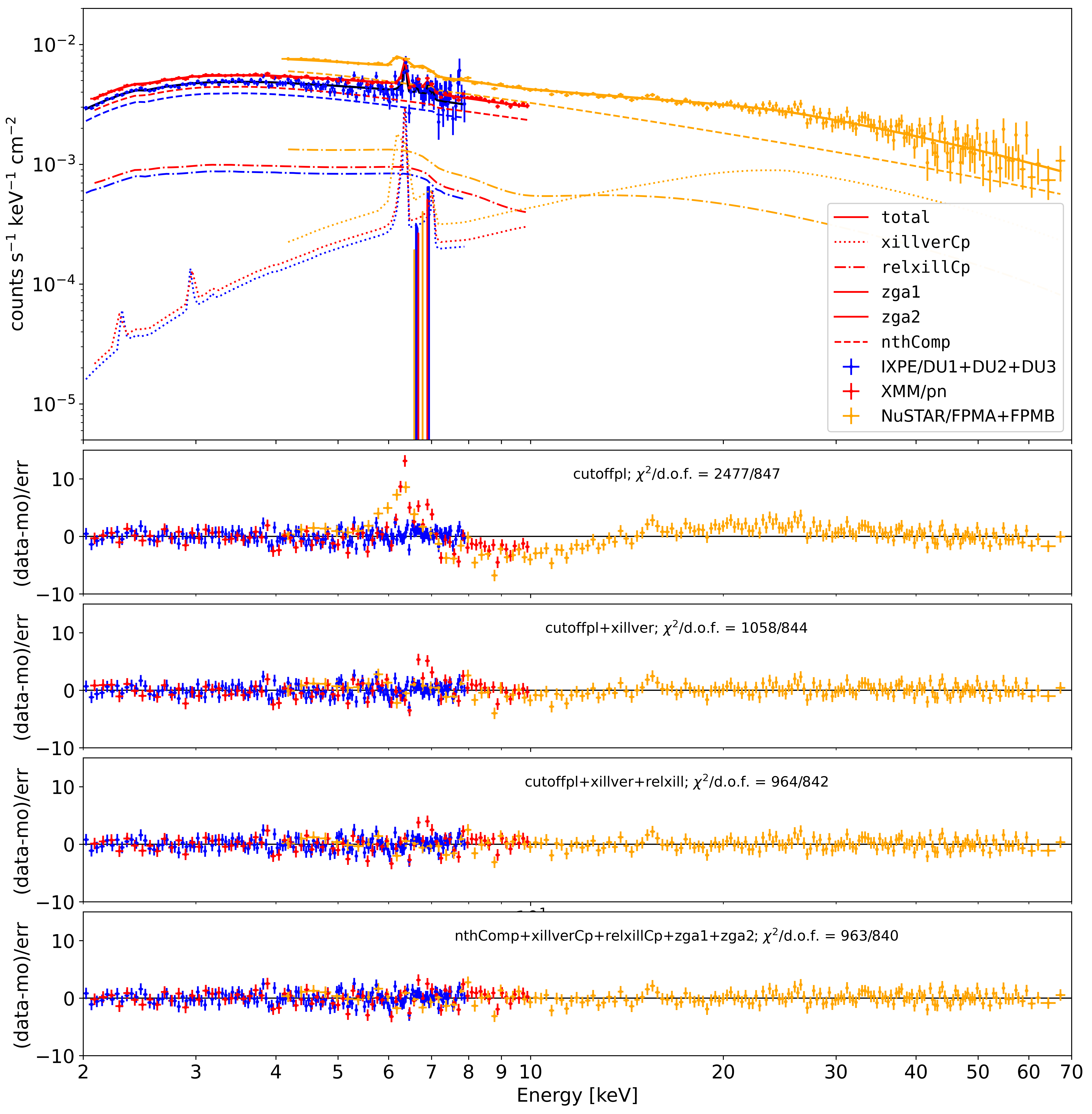}
\caption{\textit{Top panel:} \textit{NuSTAR}, \textit{XMM-Newton}, and Stokes $I$ IXPE spectra of NGC~5506, together with the best-fit model components.
\textit{Second panel:} Residuals from the broadband spectral analysis using a model that accounts only for the primary continuum (\texttt{cutoffpl}).
\textit{Third panel:} Residuals obtained after including the primary continuum and distant reflection (i.e., \texttt{cutoffpl} + \texttt{xillver}).
\textit{Fourth panel:} Residuals obtained after including the primary continuum, distant reflection, and relativistic reflection (i.e., \texttt{cutoffpl} + \texttt{xillver} + \texttt{relxill}).
\textit{Fifth panel:} Residuals obtained with the thermally Comptonized primary continuum, distant and relativistic reflection, and Gaussian emission lines (i.e., \texttt{nthComp} + \texttt{xillverCp} + \texttt{relxillCp} + \texttt{zga1} + \texttt{zga2}).
For clarity, we grouped the data from IXPE DU1, DU2, and DU3, as well as the \textit{NuSTAR} FPMA and FPMB data.}
\label{Fig:all_nopol}
\end{figure*}

We started with a baseline model composed of a simple absorbed power law with a high-energy cut off:\\\\  $\texttt{const} \times \texttt{tbabs} \times \texttt{ztbabs} \times \texttt{cutoffpl}$\\\\
Since the high energy cutoff is completely unconstrained, we fixed its value to $E_C=500$ keV, consistent with previous analyses of the hard X-ray spectrum of NGC~5506 (see e.g., \citealt{matt10.1093/mnras/stu2653}).
We also fixed the Galactic column density along the line of sight to $N_{\mathrm{H,Gal}}=4.23\times10^{20}~\mathrm{cm}^{-2}$ \citep{HI4PI2016}.
This resulted in an unsatisfactory fit ($\chi^2$/d.o.f. = 2477/847), with large residuals close to the 6.4~keV iron K$\alpha$ line and at higher energies, suggesting the presence of unmodeled reflection features (e.g., narrow emission lines and a Compton hump; see the second panel of Fig. \ref{Fig:all_nopol}).

We therefore added a \texttt{xillver} \citep{xill2013ApJ...768..146G} component to model cold reflection from the outer accretion disk and/or the torus.
We fixed the metallicity to the solar value and the logarithm of the ionization to $0$, corresponding to a neutral medium.
We also tied the spectral index and the high energy cut off to the values of the primary continuum.
Since the inclination of the cold reflector ($\theta_{\rm incl}^{\rm xill}$) was completely unconstrained, we fixed it to the standard value of $30\degree$.
Consistent with previous studies (e.g., \citealt{Marinucci_2022, taglia2023MNRAS.525.4735T}), we added a \texttt{vashift} component, which simply provides an energy shift to the data.
We used this component to verify whether the energy of the narrow Fe K$\alpha$ line was consistent with 6.4 keV in the host galaxy rest frame.
We applied this component to the \textit{XMM} and \textit{NuSTAR} datasets only, since we had already used \texttt{gain fit} for IXPE to check for calibration issues in the \textit{I} spectra.
For \textit{XMM} we obtain a velocity shift compatible with zero (suggesting that there are no deviations in these datasets), while for \textit{NuSTAR} we obtain $v_{\textnormal{FPMA}}^{\textnormal{shift}}=2.3_{-1.3}^{+0.9}\times10^3$ km s$^{-1}$ and $v_{\textnormal{FPMB}}^{\textnormal{shift}}=3.5_{-0.9}^{+1.2}\times10^3$ km s$^{-1}$, which correspond to a shift in the energy of the centroid $\Delta E_{K\alpha}=50^{+20}_{-30}$ eV for \textit{NuSTAR}/FPMA and $\Delta E_{K\alpha}=75^{+25}_{-20}$ eV for \textit{NuSTAR}/FPMB.
All of this led to a significant improvement in the fit, with $\chi^2$/d.o.f. = 1058/844.
However, some residuals were still present, suggesting the presence of unmodeled reflection features (see the third panel of Fig. \ref{Fig:all_nopol}).

For this reason, we added relativistic reflection from the inner accretion disk to the model using \texttt{relxill} \citep{xill2013ApJ...768..146G}.
We fixed the black hole spin to $0.998$, allowing the inner radius of the accretion disk ($R_{\textnormal{in}}^{\textnormal{disk}}$) to reach the lowest possible values, and fixed the metallicity to the default solar value.
We modeled the primary source of radiation that illuminated the disk without assuming a specific geometry or physical location. We tied its spectrum to that of the \texttt{cutoffpl} component, while we parametrized its radial dependence (i.e., emissivity profile) as $r^{-3}$.
The free parameters are the inclination of the disk ($\theta_{\rm incl}^{\rm rel}$), which we tied to the inclination of the \texttt{xillver} component and allowed to vary freely, $R_{\textnormal{in}}^{\textnormal{disk}}$, and the ionization ($\log{\xi}$).
This led to an improvement in the fit, with $\chi^2$/d.o.f. = 964/842.
However, some residuals were still present around 7~keV, suggesting the presence of unmodeled emission lines (see the fourth panel in Fig. \ref{Fig:all_nopol}).

For this reason, following previous observations (e.g., \citealt{bianchirefId0}), we added two narrow (i.e., $\sigma=0$) Gaussian lines using \texttt{zgauss} in \texttt{xspec}, at $6.70$ and $6.97$ keV, to model Fe~XXV He$\alpha$ and Fe~XXVI Ly$\alpha$, respectively.
This led to a marginally significant improvement in the fit, resulting in $\chi^2$/d.o.f. = 953/840.

To obtain a more physically motivated picture of the system, we replaced \texttt{cutoffpl} with a more realistic model, \texttt{nthComp} \citep{zdr1996MNRAS.283..193Z, zyc1999MNRAS.309..561Z}, which represents a thermally Comptonized continuum and allowed us to evaluate the coronal temperature. We also replaced \texttt{xillver} and \texttt{relxill} with their counterparts, which are better suited to a Comptonized continuum model than a simple power law (i.e., \texttt{xillverCp} and \texttt{relxillCp}).
Since the fit did not constrain the coronal temperature, we fixed it to a value compatible with previous estimates (i.e., $kT_e=150$ keV).
With this model, we obtain a nearly equivalent fit with $\chi^2$/d.o.f. = 963/840.
Since we considered \texttt{nthComp} to be physically more realistic than a simple cutoff power law, we adopted this as the best-fit model.
We built the final model as follows:\\\\  $\texttt{const} \times \texttt{tbabs} \times \texttt{ztbabs} \times \texttt{vashift}(\texttt{nthComp} + \\ \texttt{relxillCp}+\texttt{xillverCp}+\texttt{zga1}+\texttt{zga2})$\\\\
Using this model, we obtained a column density at the redshift of the source of $N_{H, z}=(3.01_{-0.02}^{+0.03})\times10^{22}~\mathrm{cm}^{-2}$, $R_{\textnormal{in}}^{\textnormal{disk}}=4^{+2}_{-1}$ R$_{\textnormal{G}}$, $\log{\xi}=3.4_{-0.1}^{+0.2}$, and $\Gamma=1.88^{+0.01}_{-0.02}$.
For the inclination, which we tied between \texttt{xillverCp} and \texttt{relxillCp}, we obtain only an upper limit of $\theta_{\textnormal{incl}}<37\degree$ at the $99\%$ confidence level. This value is lower than previous estimates by \cite{sun2018MNRAS.478.1900S} and compatible with our estimates obtained by comparing Monte Carlo simulations with the polarization properties derived from the IXPE data (see Sect. \ref{monksim}).
The reflection strength ($R$), defined here as the ratio between the 20--40~keV fluxes of the Compton reflection (i.e., \texttt{xillver}) and the primary component (i.e., \texttt{cutoffpl}), resulted in $R=0.6_{-0.1}^{+0.2}$.

Figure \ref{Fig:all_nopol} shows the broadband data used for this spectral analysis (top panel), along with the final model and the residuals (fifth panel).
In Tab. \ref{Tab:all_nopol}, we also show the best-fit parameters obtained from this broadband analysis using this final model.

\begin{table}[t]
\centering
\caption{Best-fitting parameters for the broadband spectral analysis of NGC~5506 using \textit{NuSTAR} (3--79~keV), \textit{XMM-Newton} (2--10~keV), and \textit{IXPE I} (2--8~keV) datasets.}
\begin{tabular}{cc}
\hline
{Parameter}  & {Best fitting value} \\
\hline
$N_{\rm H,Gal}$  [cm$^{-2}$] & $4.23\times10^{20}$*  \\
$N_{\rm H,z}$ [cm$^{-2}$] & $(3.01_{-0.02}^{+0.03})$$\times10^{22}$  \\
\hline
\multicolumn{2}{c}{\texttt{nthComp}} \\
 $\Gamma$& $1.88_{-0.02}^{+0.01}$ \\
 $kT_e$ [keV]& $150$*  \\
 $N_{\rm comp}$ &$1.8^{+0.1}_{-0.2}$$\times10^{-2}$  \\
 \hline
  \multicolumn{2}{c}{\texttt{relxillCp}} \\
  $\theta_{\textnormal{incl}}$ [deg]&$<37$\\
  $a$&0.998*\\
  $R_{\textnormal{in}}^{\textnormal{disk}}$ [$R_{\rm G}$]&$4_{-1}^{+2}$\\
  $\log{\xi}$ & $3.4_{-0.1}^{+0.2}$\\
   $N_{\rm rel}$&$9_{-1}^{+2}$$\times10^{-5}$  \\
  \hline
    \multicolumn{2}{c}{\texttt{xillverCp}} \\
  $A_{\textnormal{Fe}}$ &$1$*\\
   $N_{\rm xill}$&$2.2_{-0.2}^{+0.1}$$\times10^{-4}$  \\
  \hline
   \multicolumn{2}{c}{\texttt{zga1}} \\
  $E_1$ [keV] &$6.70$*\\
  $\sigma_1$ [keV] & $0$*\\
   $N_1$&$(9\pm3)$$\times10^{-6}$  \\
  \hline
     \multicolumn{2}{c}{\texttt{zga2}} \\
  $E_2$ [keV] &$6.97$*\\
  $\sigma_2$ [keV] & $0$*\\
   $N_2$&$(6_{-4}^{+3})$$\times10^{-6}$  \\
  \hline
  \multicolumn{2}{c}{\texttt{vashift} [km/s]}  \\
 \textit{NuSTAR}/FPMA & $2.3_{-1.3}^{+0.9}\times 10^3$  \\
  \textit{NuSTAR}/FPMB & $3.5_{-0.9}^{+1.2}\times 10^3$  \\
 \textit{XMM}  & $0$*  \\
 IXPE & $0$*  \\
  \hline
    \multicolumn{2}{c}{Cross-calibration constants}  \\
    \textit{XMM-Newton} & $1$*  \\
 IXPE/DU1 & $1.21\pm0.01$  \\
 IXPE/DU2 & $1.20\pm0.01$  \\
 IXPE/DU3 & $1.17\pm0.01$  \\
 \textit{NuSTAR}/FPMA & $1.41\pm0.01$  \\
 \textit{NuSTAR}/FPMB & $1.40\pm0.01$  \\
  \hline
     \multicolumn{2}{c}{Gain fit constants}  \\
 IXPE/DU1/slope & $1.00\pm0.01$  \\
 IXPE/DU1/offset & $0.01\pm0.02$  \\
 IXPE/DU2/slope & $0.95\pm0.01$  \\
 IXPE/DU2/offset & $0.13\pm0.02$  \\
 IXPE/DU3/slope & $0.96\pm0.01$  \\
 IXPE/DU3/offset & $0.11\pm0.02$  \\
  \hline
   \multicolumn{2}{c}{ $F_{2-10}$ [erg cm $^{-2}$s$^{-1}$]} \\ [0.4ex]
  \textit{XMM-Newton}& $5.71\times10^{-11}$\\
  \hline
 $R$ & $0.6_{-0.1}^{+0.2}$\\
  \hline
  $\chi^{2}$/dof& 963/840\\
  \hline
\end{tabular}
\tablefoot{The errors are given at the 68\% confidence level and the upper limit at the 99\% confidence level for one parameter of interest. The parameter $R$ represents the reflection strength, defined as the ratio of the 20--40~keV fluxes of the Compton reflection (i.e., \texttt{xillverCp}) and the primary component (i.e., \texttt{nthComp}). We modeled the input radiation used for illuminating the disk within \texttt{relxillCp} without assumptions about its geometry and position, while its spectrum was tied to that of \texttt{nthComp}, with a radial dependence of $r^{-3}$.  $^{(\ast)}$Parameters frozen during the fit.}
\label{Tab:all_nopol}
\end{table}

As a final test, to verify that the presence of a distant reflection component is robust and does not depend on the specific reflection model adopted, we replaced the \texttt{xillverCp} component with a \texttt{BORUS} \citep{balokovic18, balokovic19} table model. 
The latter describes reprocessing by a toroidal medium with a variable covering factor, thus providing a physically motivated geometry for the distant reflector. 
We kept the overall spectral structure and the free parameters of the other components identical to those adopted in the previous model configuration.
We left the normalization of the \texttt{BORUS} component free and tied it across the different instruments, as we did for \texttt{xillverCp}, given the simultaneous observations. 
This configuration provides a satisfactory fit ($\chi^{2}$/d.o.f. = 954/839) and yields a reflection strength of $R = 0.6_{-0.1}^{+0.2}$. This value is fully consistent with those obtained in the previous model configuration, confirming the presence of a cold reflection component.

\subsection{Spectropolarimetric analysis}
\label{5506spectropol}
Next, we included the IXPE $Q$ and $U$ spectra in the analysis by assigning a convolution polarization kernel (\texttt{polconst} in \texttt{XSPEC}) to each spectral component. 
As a reference case, we initially assumed the reflection components and the Gaussian emission lines to be unpolarized. 
This model provides a fair representation of the data, with $\chi^{2}/{\rm dof}=1143/1033$. 
Under this assumption, we attributed the entire polarized signal to the primary continuum emitted by the X-ray corona. 
We obtained only a $99\%$ confidence upper limit on its polarization degree, $\Pi<4.0\%$.
This constraint is less stringent than that obtained with the simple absorbed power-law model presented in Sect.~\ref{ixpeanalysis}. 
In the latter case, we associated the polarization with the total source flux, whereas in the present analysis we attributed it only to the primary continuum. 
Since the primary continuum contributes less flux than the total emission, the corresponding polarization fraction is characterized by a larger uncertainty.

We then allowed the polarization properties of the reflection components to vary freely.
However, this did not provide any meaningful improvement in the constraints: the polarization parameters remained largely unconstrained, regardless of whether we allowed the relativistic and distant reflection components to vary independently or tied them together.

Since the origin and relative contributions of the different reflection components to the observed polarization remain uncertain, we explored a series of physically motivated scenarios while continuing to assume that the Gaussian lines were unpolarized. 
In these tests, for the sake of simplicity, we tied the polarization properties of the \texttt{relxillCp} and \texttt{xillverCp} components together. 
In particular, we fixed their common polarization degree at $\Pi_{\rm refl}=15$, $20$, and $30\%$, values comparable to those predicted or inferred in previous studies (e.g., \citealt{cirUrsini_2023,giano2024A&A...691A..29G}), and assumed their polarization angle to be either parallel or perpendicular to that of the primary continuum.

Because the total 2--8~keV polarization degree is constrained to be lower than $3.1\%$, a reflection component polarized parallel to the primary continuum leads to more stringent upper limits on the continuum polarization than those obtained for unpolarized reflection. Specifically, we find $\Pi<1.1\%$, $\Pi<0.8\%$, and $\Pi<0.5\%$ for $\Pi_{\rm refl}=15\%$, $20\%$, and $30\%$, respectively. 
Conversely, when the polarization angles of the reflection and primary continuum components differ by $90\degree$, a more strongly polarized continuum is required to reproduce the observed upper limit. 
In this case, we obtain $\Pi=(5\pm3)\%$, $\Pi=(6\pm3)\%$, and $\Pi=(9\pm3)\%$ for $\Pi_{\rm refl}=15\%$, $20\%$, and $30\%$, respectively.
We quote all uncertainties and upper limits  in this section at the $99\%$ confidence level.

We repeated these tests after fixing the polarization angle of the primary continuum relative to the orientation of the narrow-line region bicone. 
The latter was measured through \textsc{[O~III]} imaging with the \textit{Hubble} Space Telescope by \cite{Fischer_2013} and has a position angle of approximately $22\degree$ with respect to north. 
We adopted this orientation as a proxy for the projected accretion disk axis and considered both a parallel configuration, $\Psi=22\degree$, and a perpendicular configuration, $\Psi=112\degree$.
The parallel configuration is motivated by previous IXPE observations of radio-quiet, unobscured AGNs, which suggest a possible alignment between the polarization angle of the primary continuum and the accretion-disk axis \citep{ingram2023MNRAS.525.5437I,Gianolli2023,giano2024A&A...691A..29G}. 
It is also consistent with radially extended coronal geometries, such as slab and wedge configurations, which are also supported by the Monte Carlo simulations presented in Sect.~\ref{sim}.
Conversely, a polarization angle perpendicular to the disk axis is compatible with vertically extended geometries, such as a spherical lamp-post corona or a conical outflow (Sect. \ref{monksim}).
For an unpolarized reflection component, we obtain $\Pi<3.7\%$ when we fixed the primary continuum polarization parallel to the accretion disk axis and $\Pi<1.8\%$ when we fixed it perpendicular to the axis. 
In the parallel continuum configuration, a reflection component polarized parallel to the continuum yields $\Pi<1.1\%$, $\Pi<0.8\%$, and $\Pi<0.5\%$ for $\Pi_{\rm refl}=15\%$, $20\%$, and $30\%$, respectively. 
When the reflection polarization is instead perpendicular to the continuum, we obtain $\Pi=(5\pm3)\%$, $\Pi=(6\pm3)\%$, and $\Pi=(9\pm3)\%$.
When we fixed the primary continuum polarization perpendicular to the accretion disk axis, a reflection component polarized parallel to the continuum resulted in $\Pi<0.7\%$, $\Pi<0.5\%$, and $\Pi<0.4\%$ for $\Pi_{\rm refl}=15\%$, $20\%$, and $30\%$, respectively. 
For a reflection component polarized perpendicular to the continuum, we instead obtained $\Pi<6\%$, $\Pi=(4\pm3)\%$, and $\Pi=(7\pm3)\%$ for the same three values of $\Pi_{\rm refl}$.

When we allowed the polarization angle of the primary continuum, $\Psi$, to vary freely, it remained unconstrained at the $99\%$ confidence level. 
This was also the case in the fits for which the continuum polarization degree was formally constrained rather than represented by an upper limit. 
Therefore, we could not claim a significant measurement of $\Psi$. 
This prevents a direct comparison with the expected orientation of the accretion disk axis and, consequently, does not allow us to exclude specific coronal geometries based on the polarization angle alone.

Finally, we performed additional tests assuming an unpolarized primary continuum and exploring different polarization configurations for the reflection components. 
The aim was to qualitatively reproduce the scenario adopted by \cite{Kammoun2026} for NGC~4151, in which the relatively high observed polarization is explained through a combination of an unpolarized coronal component, as may be produced by a lamp-post geometry, and a highly polarized relativistic reflection component with a polarization angle parallel to the accretion disk axis. 
Studies have previously investigated similar scenarios \citealt{dov2011ApJ...731...75D}, \citealt{pod2023MNRAS.524.3853P}, and \citealt{pod2026A&A...711A.175P}.
We first tied the polarization properties of the \texttt{relxillCp} and \texttt{xillverCp} components, obtaining $\Pi_{\rm refl}<13\%$. 
We then allowed their polarization degrees to vary independently, obtaining $\Pi_{\rm xill}<48\%$ and $\Pi_{\rm rell}<16\%$. 
Finally, following the results of \citealt{cirUrsini_2023}, we fixed the polarization angle of the cold reflection component at $112\degree$, perpendicular to the assumed accretion disk axis. 
Under this assumption, $\Pi_{\rm xill}$ remains completely unconstrained, whereas $\Pi_{\rm rell}<8\%$.

Together with the tests in which we fixed the reflection polarization degree at relatively high values, these results show that the IXPE data can, in principle, also be reproduced by a scenario involving a weakly polarized or unpolarized corona and strongly polarized reflection components. 
More generally, these tests demonstrate that the polarization properties inferred for the primary continuum depend strongly on the assumptions made about the reflected emission. 
Simultaneously constraining both contributions is therefore essential for obtaining meaningful information on the intrinsic coronal polarization and, ultimately, on the coronal geometry.
Additional information could be obtained from polarimetric observations extending into the hard X-ray band, particularly across the Compton hump, where the relative contribution of the reflected emission is expected to be significantly larger and its polarization signature could be more readily distinguished from that of the primary continuum \citep{dov2011ApJ...731...75D, pod2023MNRAS.524.3853P}. 
Such measurements could provide more direct constraints on the polarization properties of the reflection components and help disentangle their contribution from the intrinsic coronal polarization.
The current degeneracy is further highlighted by the similar $\chi^{2}$ values obtained for all the tested configurations, indicating that the available data cannot distinguish between the different scenarios. 
Nevertheless, for completeness, in the following section we present dedicated Monte Carlo simulations of the polarization properties expected for this source under different assumptions about the coronal geometry.

\section{Monte Carlo simulations}
\label{sim}
To interpret our measurements, we examined the possible shape of the X-ray corona by performing Monte Carlo simulations of the expected polarization properties (i.e., $\Pi$ and $\Psi$) of the primary continuum. 
For this purpose, we used the \texttt{MONK} code, a Monte Carlo radiative transfer tool specifically created to compute Comptonized spectra in Kerr spacetime, including all general relativistic effects \citep{Zhang2019}.
We assumed physical parameters compatible with those obtained for this source (see Sect. \ref{Introduction}): a maximally spinning black hole ($a=0.998$) with a mass $M_{BH}=10^6$ $M_{\odot}$, an accretion rate $\dot{m}=0.12$, a coronal temperature $kT_{\rm e}=100$ keV, and a primary continuum spectral index $\Gamma=1.9$.
We assumed that the accretion disk extends down to the innermost stable circular orbit (ISCO; $1.24$ R$_{\textnormal{G}}$ for $a=0.998$) to $1000$ R$_{\textnormal{G}}$, emits optical and UV radiation, and is polarized as expected for a pure-scattering, plane-parallel, semi-infinite atmosphere \citealt{chandra1960ratr.book.....C}.

We tested the coronal models described in Sect. \ref{Introduction}: a spherical lamp-post, a conical outflow, a slab corona, and a wedge corona. 
The lamp-post model has a height above the accretion disk $H=5$ R$_{\textnormal{G}}$ and a radius $R_C=2$ R$_{\textnormal{G}}$, with a Thomson optical depth ($\tau$)  of $0.82$ required to reproduce $\Gamma=1.9$.
The cone model ($\tau=0.58$) has a height above the disk $d=3$ R$_{\textnormal{G}}$, a thickness $t=10$ R$_{\textnormal{G}}$, an opening angle $\theta=30\degree$, and an outflowing velocity $v=0.3c$.
The slab ($\tau=0.19$) model reflects the inner and outer radii of the disk and has a height $h=1$ R$_{\textnormal{G}}$.
Finally, the wedge model ($\tau=1.65$) has an opening angle from the accretion disc $\alpha=30\degree$, an inner radius that coincides with the ISCO, and an outer radius of $25$ R$_{\textnormal{G}}$.

As expected, the slab and wedge geometries result in polarization directions parallel to the accretion disk axis, while the lamp-post and cone geometries result in directions perpendicular to it.
For the polarization fraction, the slab resulted in $\Pi$ values of up to $8\%$, the wedge up to $10\%$, the cone up to $2\%$, and the lamp-post, being the most symmetrical geometry, up to $1.5\%$.
Since all these models are axisymmetric, they produce polarization fractions ranging from $0\%$ to their maximum value as the system is viewed from a face-on to edge-on.

\begin{figure}[h!]
\centering
\includegraphics[width=9cm] {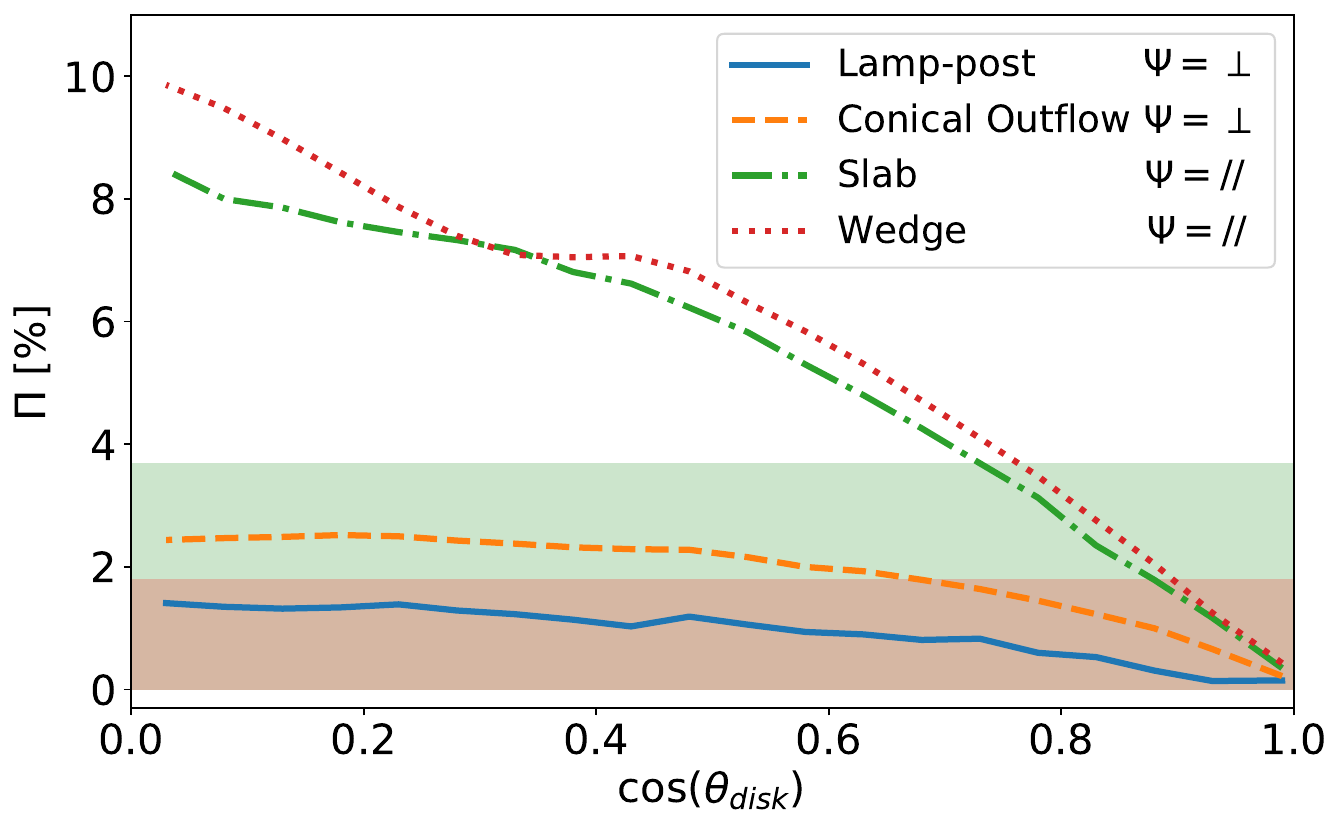}
\caption{Results of Monte Carlo simulations performed with the \texttt{MONK} code. The polarization fraction ($\Pi$) as a function of the cosine of the inclination of the disk ($\cos{\theta_{\textnormal{disk}}}$) is shown for various coronal geometries: spherical lamp-post (blue), conical outflow (orange), slab (green) and wedge (red). The green and red regions represent the constraints on $\Pi$ obtained from a model-dependent analysis ($\Pi<3.7\%$ for $\Psi$ parallel to the accretion disk axis: green region; $\Pi<1.8\%$ for $\Psi$ perpendicular: red region; see Sect. \ref{5506spectropol}). The // and $\perp$ symbols indicate the direction relative to the accretion disk axis.}
\label{monksim}
\end{figure}

Figure \ref{monksim} shows the results of the Monte Carlo simulations as a function of the cosine of the disk inclination.
The green and red regions represent the constraints on the primary continuum polarization fraction obtained from the model-dependent analysis in Sect. \ref{5506spectropol}. We compared the simulation outcomes with the upper limits obtained by assuming a zero polarization fraction for the reflection component and a polarization angle for the primary continuum that is either parallel ($\Pi<3.7\%$) or perpendicular ($\Pi<1.8\%$) to the accretion disk axis ($\sim22\degree$).
While the assumption on $\Pi_{\rm refl}$ is conservative, we consider the assumptions on $\Psi$ to be the most realistic, since all tested coronal models result in coronal radiation polarized in these directions (i.e., the slab and wedge parallel, and the lamp-post and cone perpendicular to, the accretion disk axis). 
When we compared the simulations with the spectropolarimetric measurements,  the lamp-post model provides no constraints on the disk inclination angle, whereas the conical model disfavors viewing angles larger than $45\degree$. Adopting either the slab or wedge geometry (as suggested by previous observations of similar sources) is inconsistent with angles larger than $40\degree$.
These results are in contrast with previous analyses, which have yielded higher inclinations. 
A relativistic reflection analysis by\cite{sun2018MNRAS.478.1900S} gives a moderate disk inclination ($40\degree< \theta_{\rm disk} < 50\degree$), while \cite{Fischer_2013} found a much higher inclination ($\theta_{\rm disk}\sim80\degree$) using [O~III] imaging.
In contrast, these results are fully consistent with those obtained from the broadband spectral analysis presented in Sect. \ref{spectralanalysis}, which constrained $\theta_{\rm{incl}}$ to be below $37\degree$ at the $99\%$ confidence level.

\section{Conclusions}
\label{concl}
In conclusion, the observation of the Compton-thin Seyfert galaxy NGC~5506 with IXPE provides new constraints on its X-ray polarization properties. 
The net exposure of 975~ks, partially simultaneous with \textit{NuSTAR} and \textit{XMM-Newton}, allowed us to perform a model-independent analysis, yielding a polarization degree of $\Pi=(1.0\pm0.8)\%$ and a $99\%$ confidence upper limit of $\Pi<3.1\%$. 
We performed a subsequent spectropolarimetric analysis, which combined data from the three observatories. This revealed the complexity of the X-ray emission, including significant contributions from both relativistic and distant cold reflection. 
The relatively strong contribution of these components led to substantially different constraints on the polarization of the primary continuum, depending on the assumptions about the polarization properties of the reflected emission.

We also explored a scenario similar to that proposed by \cite{Kammoun2026} for NGC~4151, in which the coronal emission is unpolarized or only weakly polarized, while a significant fraction of the observed polarization originates from a highly polarized relativistic reflection component. 
Our tests show that such a configuration can, in principle, reproduce the IXPE data of NGC~5506. 
However, the polarization properties of the individual reflection components remain poorly constrained, and the comparable fit statistics obtained for the different configurations prevent us from favoring this scenario over models in which the polarization is primarily associated with the coronal continuum.

Furthermore, our Monte Carlo simulations with the radiative-transfer code \texttt{MONK} highlight the difficulty of drawing firm conclusions about the coronal geometry. 
The weak constraints on both $\Pi$ and $\Psi$, together with their dependence on the assumed reflection polarization, prevent us from robustly excluding specific coronal configurations. Nevertheless, assuming a slab or wedge corona, as suggested for similar sources, our simulations disfavor inclination angles larger than $40\degree$.
This result contrasts with some previous estimates but is consistent with the inclination inferred from the broadband spectral analysis presented in this work.

\begin{acknowledgements}
The Imaging X-ray Polarimetry Explorer (IXPE) is a joint US and Italian mission. The US contribution is supported by the National Aeronautics and Space Administration (NASA) and led and managed by its Marshall Space Flight Center (MSFC), with industry partner Ball Aerospace (contract NNM15AA18C). DT, SB, GM, and FU acknowledge financial support by the Italian Space Agency (Agenzia Spaziale Italiana, ASI) through the contract ASI-INAF-2022-19-HH.0. VEG acknowledges funding under NASA contract 80NSSC24K1403. AG was supported by an appointment to the NASA Post-doctoral Program at the Marshall Space Flight Center (MSFC), administered by Oak Ridge Associated Universities under contract with NASA. AT acknowledges financial support from the Bando Ricerca Fondamentale INAF 2022 Large Grant 'Toward a holistic view of the Titans: multi-band observations of $z>6$ QSOs powered by greedy supermassive black holes' and from the Bando Ricerca Fondamentale INAF 2024 Large Grant 'The DEepest study of LUminous QSOs in X-ray at z=2-7'.
\end{acknowledgements}

\bibliographystyle{aa}
\bibliography{Bibliography}

\end{document}